\documentclass[twocolumn,letterpaper,aps,prc,longbibliography,superscriptaddress,showpacs,floatfix,nofootinbib]{revtex4-1}

\usepackage{graphicx}	
\usepackage{xspace}     
\usepackage{amsmath}
\usepackage[usenames,dvipsnames]{color}

\newcommand{\sqsn}{\mbox{$\sqrt{s_{_{NN}}}$}\xspace}

\def\lsim{\raise0.3ex\hbox{$<$\kern-0.75em\raise-1.1ex\hbox{$\sim$}}}
\def\gsim{\raise0.3ex\hbox{$>$\kern-0.75em\raise-1.1ex\hbox{$\sim$}}}
\def\mean#1{\left<#1\right>}
\def\Journal#1#2#3#4{{#1}{\bf #2}, #3 (#4)}

\def\JPCS{{J. Phys: Conf. Series\ }}

\def\NPA{{Nucl. Phys. A}}
\def\NPB{{Nucl. Phys. B}}
\def\PLB{{Phys. Lett. B}}
\def\PLC{Phys. Repts.\ }

\def\PRL{Phys. Rev. Lett.\ }
\def\PRD{{Phys. Rev. D}}
\def\PRC{{Phys. Rev. C}}

\def\la{\left< }
\def\ra{\right> }
\def\jt#1{\ensuremath{j_{T\rm #1}}}
\def\meankv#1{\ensuremath{\la#1^2\ra}}
\def\rms#1{\meankv{#1}}

\begin{document}
\title{Measurement of $\hat{q}$ in Relativistic Heavy Ion Collisions using di-hadron correlations.}

\author{M.~J.~Tannenbaum\\ {\it Brookhaven National Laboratory, Upton, NY 11973 USA}}

\date{\today}

\begin{abstract}
The propagation of partons from hard scattering through the Quark Gluon Plasma produced in A$+$A collisions at RHIC and the LHC is represented in theoretical analyses by the transport coefficient $\hat{q}$ and predicted to cause both energy loss of the outgoing partons,  observed as suppression of particles or jets with large transverse momentum $p_T$, and broadening of the azimuthal correlations of the outgoing di-jets or di-hadrons from the outgoing parton-pair, which has not been observed. The widths of azimuthal correlations of di-hadrons with the same trigger particle $p_{Tt}$ and associated $p_{Ta}$ transverse momenta in p$+$p and Au$+$Au are so-far statistically indistinguishable as shown in recent as well as older di-hadron measurements and also with jet-hadron and hadron-jet measurements.   The azimuthal width of the di-hadron correlations in p$+$p collisions, beyond the fragmentation transverse momentum, $j_T$, is dominated by $k_T$, the so-called intrinsic transverse momentum of a parton in a nucleon, which can be measured. The broadening should produce a larger $k_T$ in A$+$A than in p$+$p collisions. The present work introduces the observation that the $k_T$ measured in p$+$p collisions for di-hadrons with $p_{Tt}$ and $p_{Ta}$ must be reduced to compensate for the energy loss of both the trigger and away parent partons when comparing to the $k_T$ measured with the same di-hadron $p_{Tt}$ and $p_{Ta}$ in Au$+$Au collisions. This idea is applied to a recent STAR di-hadron measurement, with result $\mean{\hat{q}L}=2.1\pm 0.6$ GeV$^2$. This is more precise but in agreement with a theoretical calculation of $\mean{\hat{q}L}=14^{+42}_{-14}$ GeV$^2$ using the same data. Assuming a length $\mean{L}\approx 7$ fm for central Au$+$Au collisions the present result gives $\hat{q}\approx 0.30\pm 0.09$ GeV$^2$/fm, in fair agreement with the JET collaboration result from single hadron suppression of $\hat{q}\approx 1.2\pm 0.3$ GeV$^2$/fm at an initial time \mbox{$\tau_0=0.6$ fm/c} in Au$+$Au collisions at \sqsn=200 GeV.

\end{abstract}

	
\maketitle


\section{Introduction}
In the original BDMPSZ formalism~\cite{BDMPSNPB484,BSZARNPS50}, the energy loss of an outgoing parton, $-dE/dx$,  
per unit length ($x$) of a medium with total length $L$, is proportional to the 4-momentum transfer-squared, $q^2$, and takes the form:
\begin{equation}
{-dE \over dx }\simeq \alpha_s \langle{q^2(L)}\rangle=\alpha_s\, \mu^2\, L/\lambda_{\rm mfp} 
=\alpha_s\, \hat{q}\, L\qquad ,   \label{eq:Eloss}
\end {equation}
where $\mu$, is the mean momentum transfer per collision, and the transport coefficient $\hat{q}=\mu^2/\lambda_{\rm mfp}$ is the 4-momentum-transfer-squared to the medium per mean free path, $\lambda_{\rm mfp}$.  
Additionally, the accumulated momentum-squared, $\mean{p^2_{\perp W}}$ transverse to a parton traversing a length $L$ in the medium  is well approximated by~\cite{BDMPSNPB484} $\mean{p^2_{\perp W}}\approx\langle{q^2(L)}\rangle=\hat{q}\, L$. This results in a direct snd simple relationship between  the parton energy loss (Eq.~\ref{eq:Eloss}) and the di-jet azimuthal broadening, $\mean{p^2_{\perp W}}/2$, because only one of the components of the accumulated momentum transverse to the outgoing parton is in the scattering plane, the other being along the beam axis for mid-rapidity di-jets. 

It has long been established~\cite{CCHK77} that even in p$+$p collisions, or in the initial hard-scattered parton pair in A$+$A collisions, the mid-rapidity di-jets from hard-scattering are not back-to-back in azimuth but are acollinear with a net transverse momentum, $\mean{p^2_T}_{\rm pair}=2\mean{k^2_T}$, where $\mean{k_T}$ is the average `intrinsic' transverse momentum of a quark or gluon in a nucleon as defined by Feynman, Field and Fox~\cite{FFF}. Again, only the component of $\mean{p^2_T}_{\rm pair}$ perpendicular to the di-jet axis leads to acoplanarity. Thus in an A$+$A collision the relationship in Eq.~\ref{eq:qhatsolution} should hold:
\begin{equation}
\mean{\hat{q} L}/2=\mean{k_{T}^2}_{AA}-\mean{k{'}_{T}^2}_{pp}
 \label{eq:qhatsolution}
\end{equation} 
for azimuthal correlations of a trigger particle with $p_{Tt}$ and away-side particles with $p_{Ta}$. It is important to note the $'$ in $\mean{k{'}_{T}^2}_{pp}$, introduced here,  which indicates that the $k_T$ measured in p$+$p collisions for di-hadrons with $p_{Tt}$ and $p_{Ta}$ must be reduced to compensate for the energy loss of both the trigger and away parent partons when comparing to the $k_T$ calculated with the same di-hadron $p_{Tt}$ and $p_{Ta}$ in Au$+$Au collisions.  

Many experiments at RHIC, including recent experiments with di-hadron~\cite{STARPLB760}, jet-hadron~\cite{STARPRL112} and di-jet~\cite{JacobsNPA956} azimuthal correlations have searched for azimuthal broadening in Au$+$Au collisions compared to p$+$p collisions but have not found a significant difference in the azimuthal angular Gaussian width of the away-peaks.  Here we shall reexamine the STAR  di-hadron measurement~\cite{STARPLB760}  in terms of the out of plane component, $p_{\rm out}$ rather than the azimuthal angular width, taking account of the energy lost by the original parton-pair in Au$+$Au collisions when comparing to the p$+$p measurement.

\section{How information about the initial partons can be derived from two-particle correlations.} 
\begin{figure}[h]
\includegraphics[width=\linewidth]{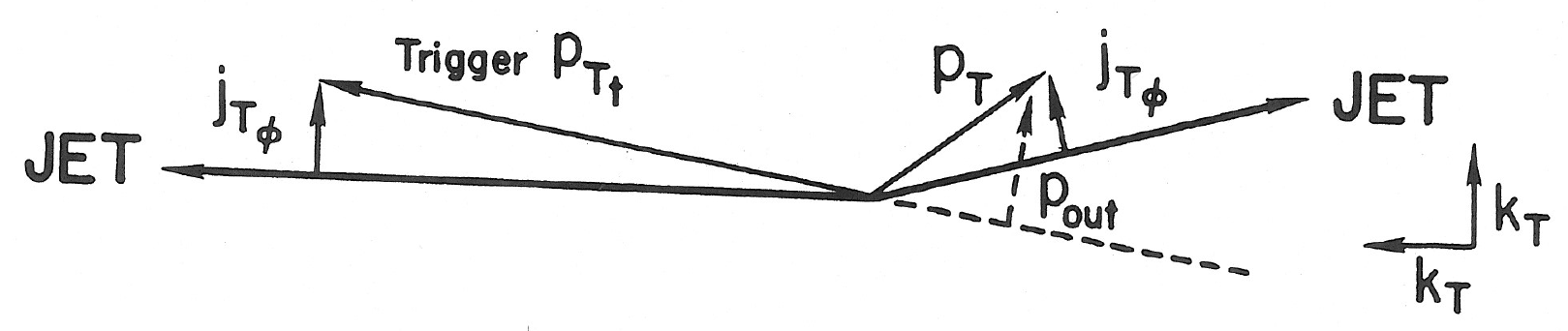}
\caption[]{Azimuthal projection of di-jet with trigger particle $p_{Tt}$ and associated away-side particle $p_{Ta}$, and the azimuthal components $j_{T\phi}$ of the fragmentation transverse momentum. The initial state $k_T$ of a parton in each nucleon is  shown schematically: one vertical which gives an azimuthal decorrelation of the jets and one horizontal which changes the transverse momentum of the trigger jet.} 
\label{fig:mjt-poutxe}
\end{figure}

We shall calculate $\rms{k_T}$ from p$+$p and Au$+$Au di-hadron measurements with the same trigger particle transverse momentum, ${p}_{Tt}$, away-side ${p}_{Ta}$ and $x_h=p_{Ta}/p_{Tt}$. The di-hadrons are assumed to be fragments of jets with transverse momenta $\hat{p}_{Tt}$ and $\hat{p}_{Ta}$ with ratio $\hat{x}_h=\hat{p}_{Ta}/\hat{p}_{Tt}$, where ${z_t}\simeq p_{Tt}/\hat{p}_{Tt}$ is the fragmentation variable, the fraction of momentum of the trigger particle in the trigger jet, and $j_T$ is the jet fragmentation transverse momentum.  The standard equation at RHIC comes from PHENIX~\cite{ppg029}, which we write in a slightly different form in Eq.~\ref{eq:kTformula2}:
  \begin{equation}
\sqrt{\mean{k^2_T}}=\frac{\hat{x}_h}{\mean{z_t}}\sqrt{\frac{ \mean{p^2_{\rm out}}-(1+{x_h^2})\rms{j_T}/2}{x_h^2}}  \qquad .
\label{eq:kTformula2}
\end{equation}   
Here $p_{\rm out}\equiv p_{Ta} \sin\Delta\phi$ (see Fig.~\ref{fig:mjt-poutxe}) and we have taken $\rms{\jt{a\phi}}=\rms{\jt{t\phi}}=\rms{j_T}/{2}$. The variable $x_h$ (which STAR calls $z_T$) is used as an approximation of the variable $x_E=x_h\cos\phi$ of the original terminology from the CERN ISR where $k_T$ was discovered and measured 40 years ago~\cite{Darriulat76,CCHK77,FFF,CCORPLB97}.  

\subsection{Bjorken parent-child relation and `trigger-bias'~\cite{JacobPLC48}} \label{sec:BJJL}
If the fragmentation function of the jet is a function only of the fragmentation variable $z$ and not of the jet $\hat{p}$, then the single particle cross section has the same power law 
shape, $d^3\sigma/2\pi p_T dp_T dy\propto p_T^{-n}$, as the parent jet  cross section. 

Furthermore, large values of $\mean{z_t}=p_{Tt}/\hat{p}_{Tt}$ dominate the single-particle cross section (e.g. $\pi^0$) used as the trigger for the di-hadron (e.g. $\pi^0$-h) measurement. This is called trigger-bias but is valid also for the simple single-particle  measurements. Calculations of $\mean{z_t}$ vs. $p_{Tt}$ for $\pi^0$ at $\sqsn=200$ GeV are given in Ref.~\cite{ppg089}.

\subsection{The energy loss of the trigger jet from p$+$p to Au$+$Au can be measured.} \label{sec:shiftmeasured}
\begin{figure}[h]
\includegraphics[width=\linewidth]{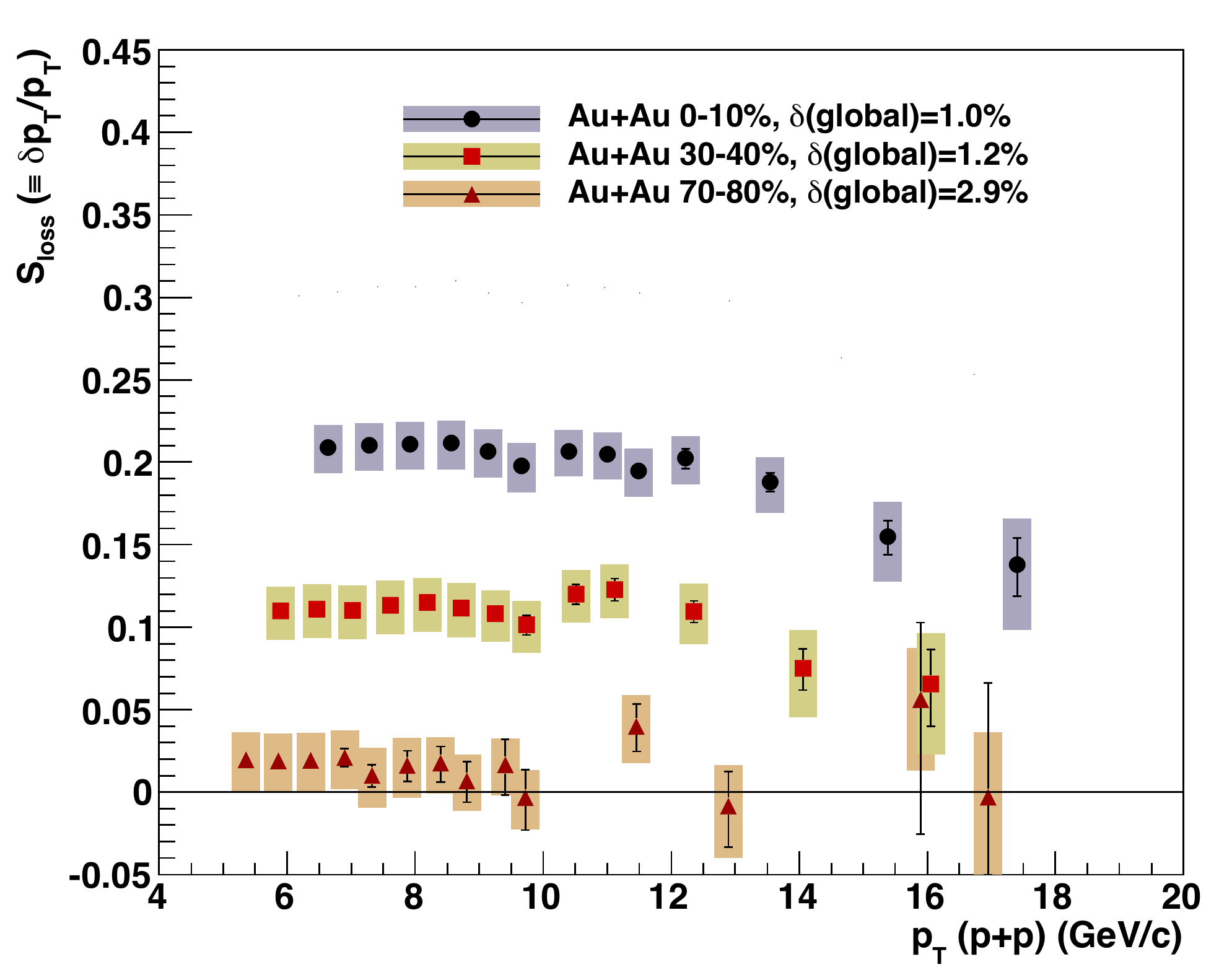}
\caption[]{$p_{T}^{pp}$ dependence of $\delta p_T/p_T^{pp}$ of $\pi^0$ from p$+$p to Au$+$Au for the
centralities indicated at \sqsn=200 GeV~\cite{ppg133}.} 
\label{fig:Sloss}
\end{figure}

At RHIC, in p$+$p and Au$+$Au collisions as a function of centrality the $\pi^0$ $p_{T}$ spectra with \mbox{$5 <p_T \ \lsim \ 20$ GeV/c} all follow the same power law with $n\approx 8.10\pm0.05$~\cite{ppg080}. From the Bj parent-child relation, the energy loss of the trigger jet is found by measuring $\delta p_T/p_T^{pp}$, the shift in the $\pi^0$ spectra in Au$+$Au at a given $p_{T}$ from the $\mean{T_{AA}}$ corrected p$+$p cross section  (Fig.~\ref{fig:Sloss})~\cite{ppg133}. The small dropoff of $\delta p_T/p_T^{pp}$ for $p_T\geq 14$ GeV/c indicates a small increase of $n$ with increasing $p_T$.  

It is important to note that the same value of $n$ for the $\pi^0$ spectra in p$+$p and Au$+$Au collisions implies the same value of $n$ for the original parton in p$+$p and the one that has lost energy in Au$+$Au. However $\mean{z_t}$ for p$+$p and Au$+$Au measurements may differ slightly because the maximum possible parton energy $\sqsn/2$ is reduced by the energy loss. The effect on $\mean{z_t}$ from p$+$p to Au$+$Au was estimated by increasing $p_{Tt}$ in the calculation of $\mean{z_t}$ in p$+$p collisions~\cite{ppg089} by the largest $\delta p_T/p_T^{pp}=0.20$ for centrality 0-10\% (Fig.\ref{fig:Sloss})  with result for $p_{Tt}=7.8$ GeV/c, $\mean{z_t}=0.63\pm 0.07$, and for $p_{Tt}=7.8/0.80=8.78$ GeV/c, $\mean{z_t}=0.66\pm 0.06$. Since the difference for the largest $\delta p_T/p_T^{pp}=0.20$ is considerably less than the error in the calculation, we shall use the measured or calculated $\mean{z_t}$ in p$+$p also for Au$+$Au with the same $p_{Tt}$.  

\begin{figure*}[!ht]\vspace*{0.0pc}
\includegraphics[width=0.49\linewidth]{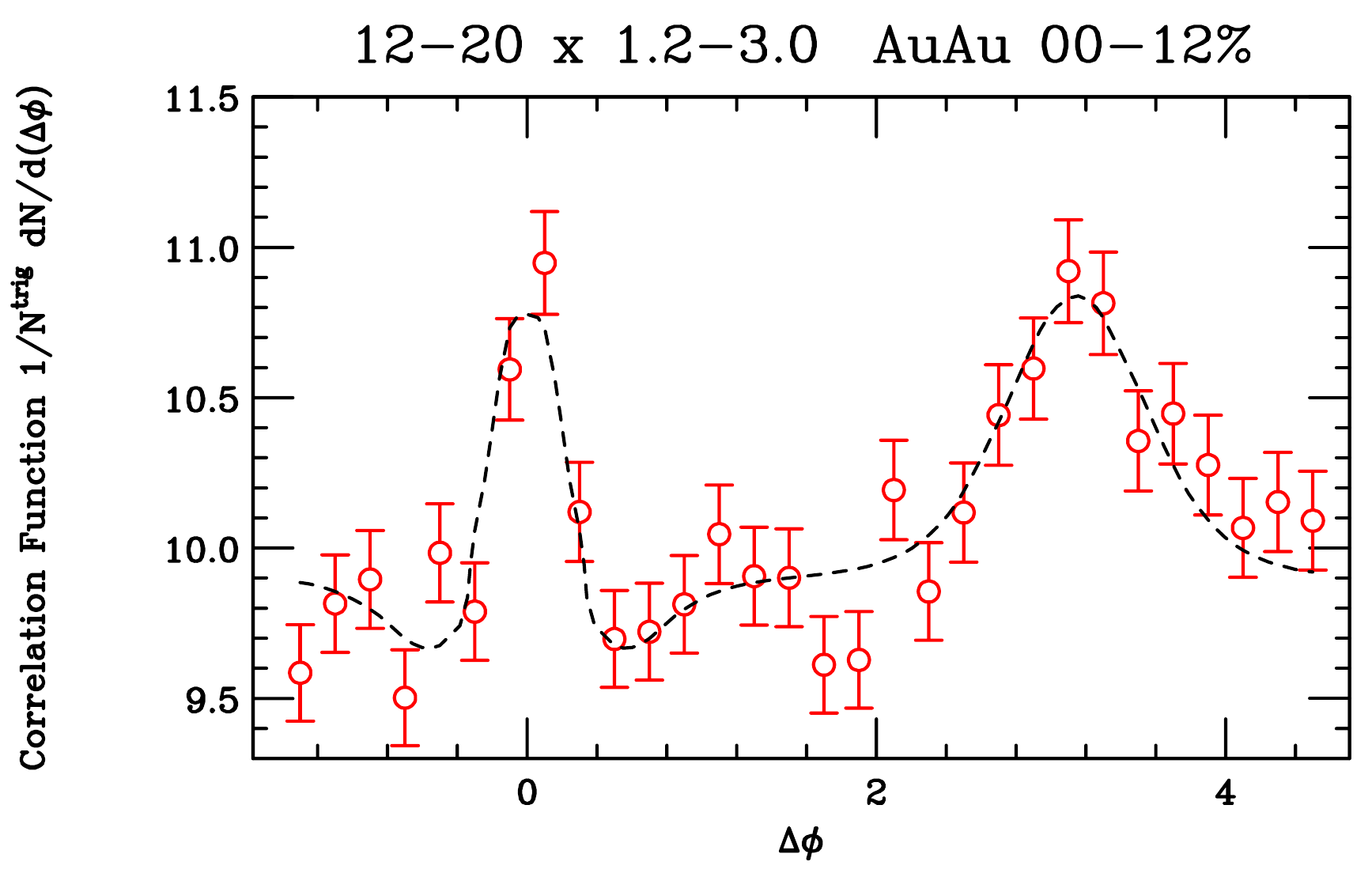}
\includegraphics[width=0.49\linewidth]{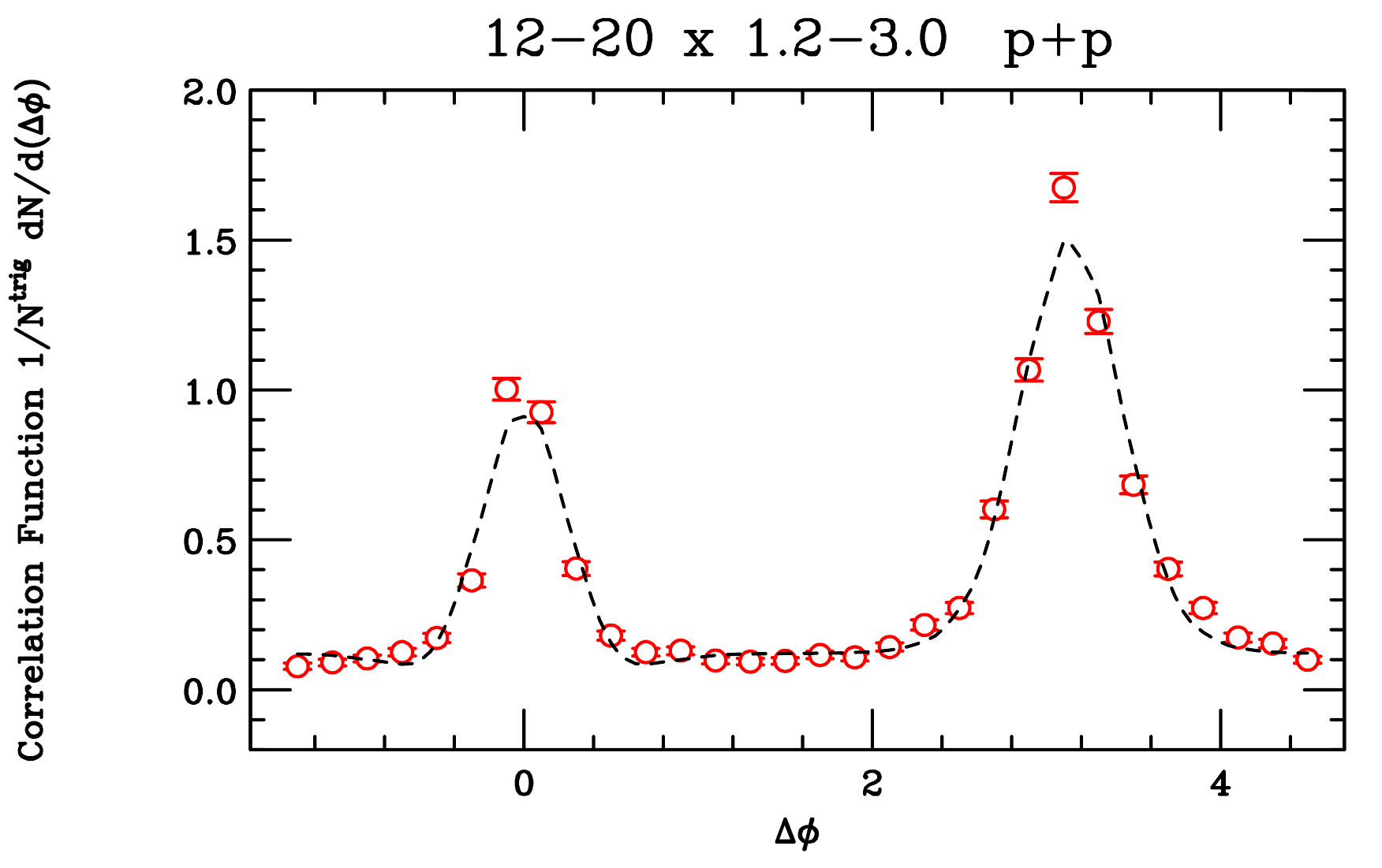}\\
\includegraphics[width=0.49\linewidth]{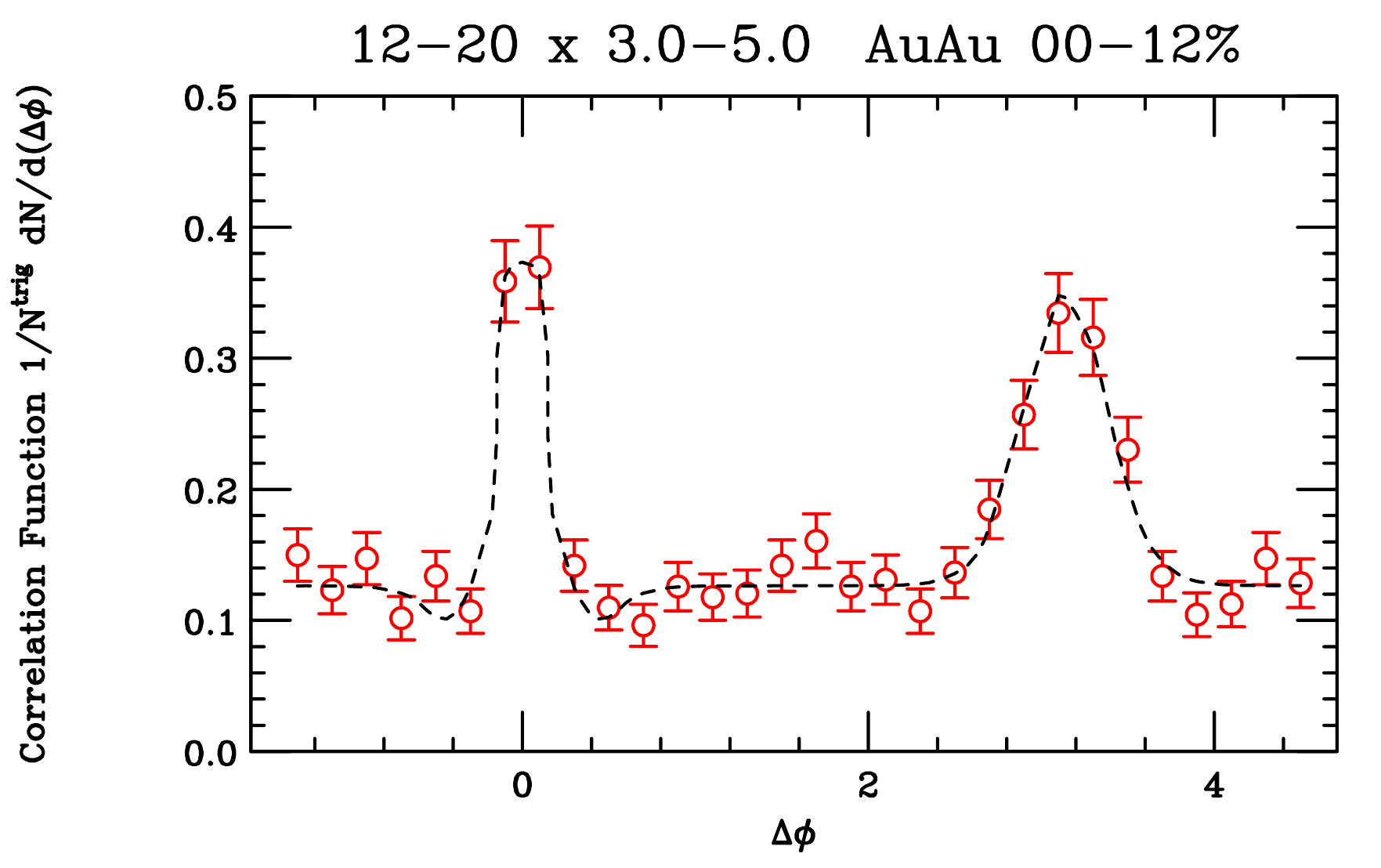}
\includegraphics[width=0.49\linewidth]{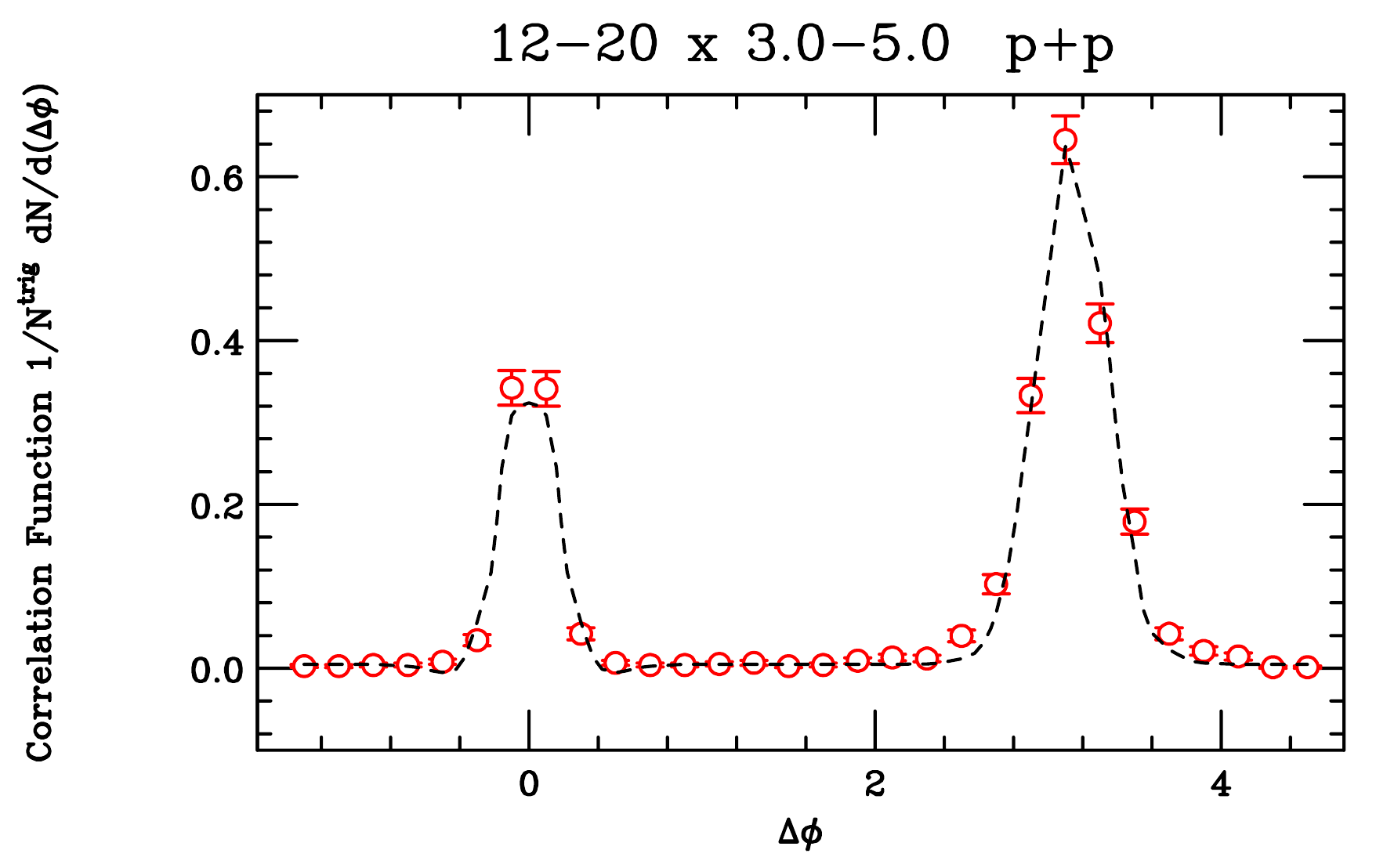}
\caption[]{Fits to STAR $\pi^0$-h correlation functions for $12<p_{Tt}<20$ GeV/c~\cite{STARPLB760} measured in central (0-12\%) Au$+$Au collisions (left) and p$+$p collisions (right): (top) $1.2<p_{Ta}<3$ GeV/c; (bottom) $3<p_{Ta}<5$ GeV/c.}  
\label{fig:MJTSTARcfnfit}
\end{figure*}\vspace*{-1.0pc}
\subsection{The away particles from a hadron-trigger do not measure the fragmentation function~\cite{ppg029}}
It was generally assumed, as implied by Feynman, Field and Fox in 1977~\cite{FFF}, that the $p_{Ta}$ (or $x_E$, or $x_h$) distribution of away-side hadrons from a single hadron trigger with $p_{Tt}$, corrected for $\mean{z_t}$, would be the same as that from a jet-trigger and would measure the away-jet fragmentation function as it does for direct photon triggers~\cite{ppg095}. However, attempts to try this at RHIC led to the discovery~\cite{ppg029} that the $x_E$ distribution does not measure the fragmentation function. The good news was that it measured the ratio of the away jet to the trigger jet transverse momenta, $\hat{x}_h=\hat{p}_{Ta}/\hat{p}_{Tt}$, Eq.~\ref{eq:condxe2N}
    \begin{equation}
\left.{dP_{\pi} \over dx_E}\right|_{p_{T_t}}  = {N\,(n-1)}{1\over\hat{x}_h} {1\over {(1+ {x_E \over{\hat{x}_h}})^{n}}} \,  
\qquad ,  
\label{eq:condxe2N}
\end{equation}   
with the value of $n=8.10$ ($\pm0.05$) fixed as determined in  Ref.~\cite{ppg080}, where $n$ is the power-law of the inclusive $\pi^0$ spectrum and is observed to be the same in p$+$p and Au+Au collisions in the $p_{T_t}$ range of interest as noted in section~\ref{sec:shiftmeasured} above.

\section{How to apply this information to find $\hat{q}$ from p$+$p and Au$+$Au di-hadron measurements}
A recent STAR $\pi^0$+h di-hadron measurement in p$+$p and Au$+$Au collision at \sqsn=200 GeV~\cite{STARPLB760} is used to measure  $\mean{\hat{q}L}$ by calculating $k_T$ in each case as in Eq.~\ref{eq:qhatsolution}.
For a di-jet produced in a hard scattering, the initial $\hat{p}_{Tt}$ and $\hat{p}_{Ta}$ will both be reduced by energy loss in the medium to become $\hat{p}{'}_{Tt}$ and $\hat{p}{'}_{Ta}$ that will be measured by the di-hadron correlations with $p_{Tt}$ and $p_{Ta}$ in Au$+$Au collisions. As both jets from the initial di-jet lose energy in the medium, the azimuthal angle between the di-jets  from the $\mean{k_{T}^2}$ in the original collision should not change unless the medium induces multiple scattering from $\hat{q}$. Thus, without $\hat{q}$ and assuming the same fragmentation transverse momentum $\rms{j_T}$ in the original jets and those that have lost energy, the   $p_{\rm out}$ between the away hadron with $p_{Ta}$ and the trigger hadron with  $p_{Tt}$ will not change (Fig.~\ref{fig:mjt-poutxe}), but the $\mean{k{'}_{T}^2}$ will be reduced according to Eq.~\ref{eq:kTformula2} because the ratio of the away to the trigger jets $\hat{x}{'}_h=\hat{p}{'}_{Ta}/\hat{p}{'}_{Tt}$ will be reduced. Thus the calculation of $k{'}_T$ from the di-hadron p$+$p measurement to compare with Au$+$Au measurement with the same di-hadron trigger $p_{Tt}$ and $p_{Ta}$ must use the values of $\hat{x}_h$, and $\mean{z_T}$ from the Au$+$Au measurement to compensate for the energy lost by the original dijet in p$+$p collisions.
\begin{figure*}[!th]
\includegraphics[width=0.49\linewidth]{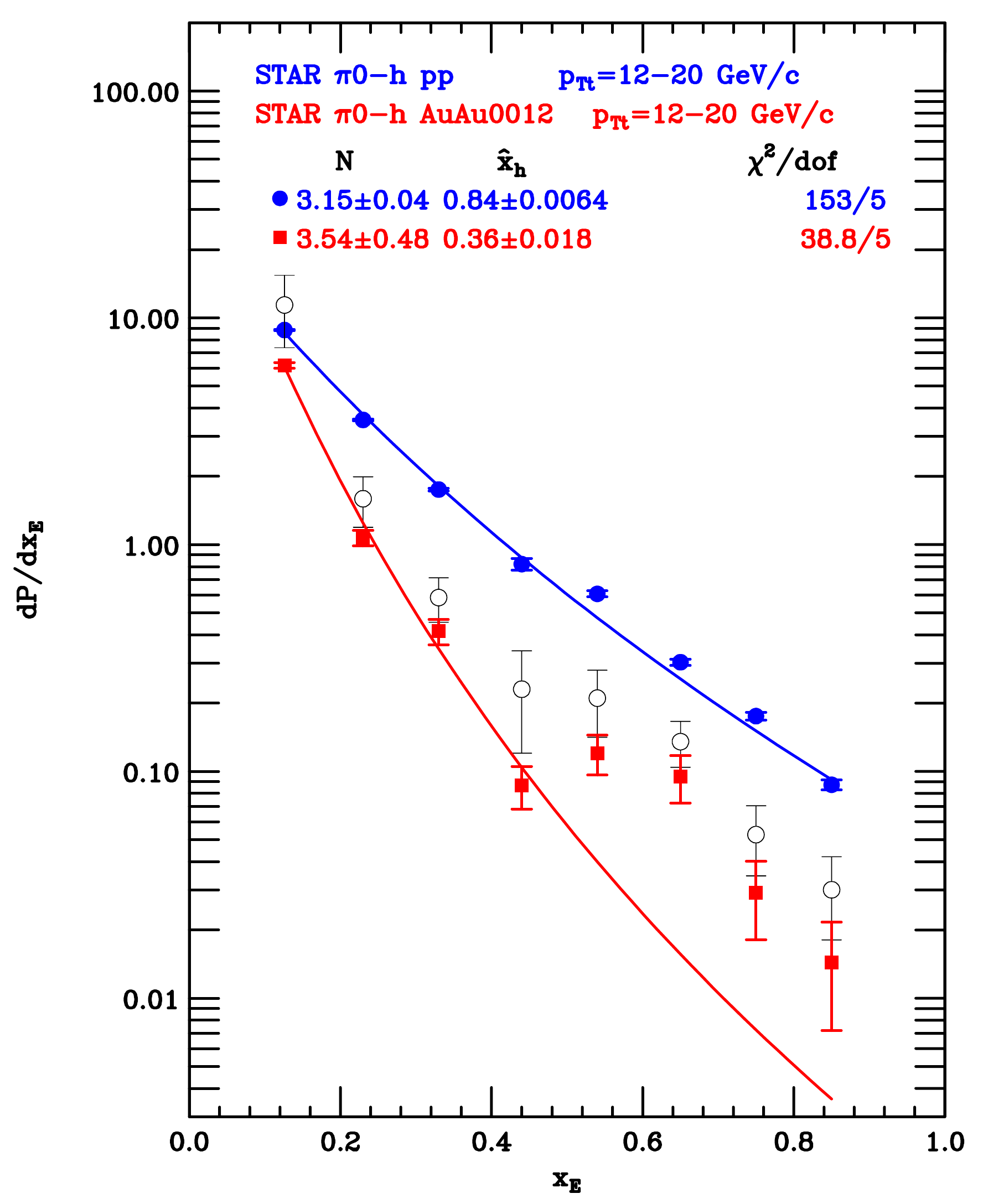}
\includegraphics[width=0.48\linewidth]{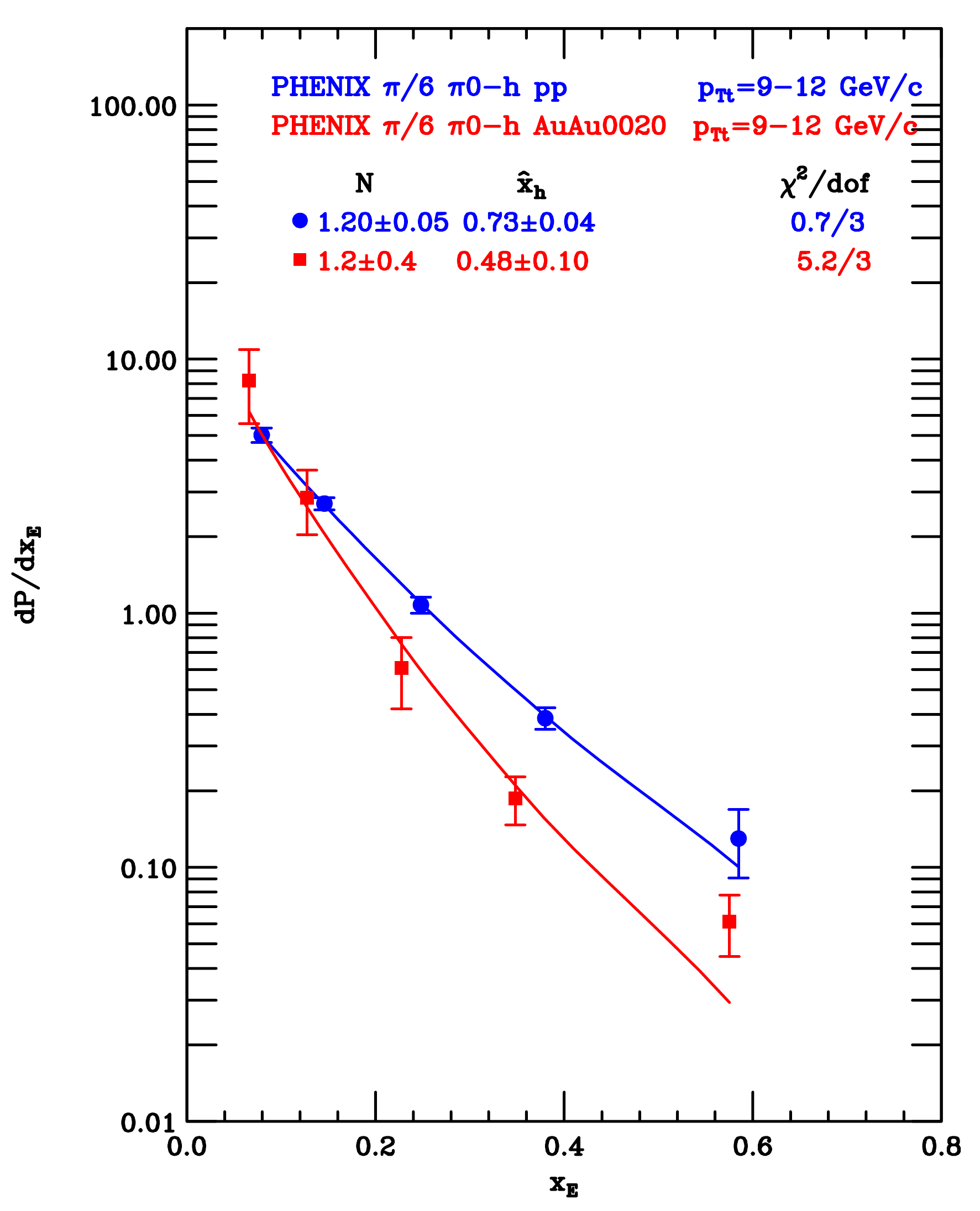}
\caption[]{a) (left) Fits of Eq.~\ref{eq:condxe2N} to the STAR away-side $z_T$ distributions~\cite{STARPLB760} in  Au$+$Au 0-12\% centrality and p$+$p for $12<p_{Tt}<20$ GeV/c. The Au$+$Au curve is a fit with $\hat{x}^{AA}_h=0.36\pm 0.05$ with error corrected by $\sqrt{\chi^2/\rm{dof}}$. The points with the open circles are the $y_i$ and systematic errors $\sigma_{b_i}$ of the data points while the filled points are $y_i+{\epsilon_b}\sigma_{b_i}$ with errors $\tilde{\sigma}_i$ and $\epsilon_b=-1.3\pm0.5$.  b) (right) Fits to PHENIX $x_E$ distributions~\cite{ppg106,MJTGrenoble} in p$+$p and Au$+$Au for $\pi^0$-h correlations with $9\leq p_{Tt}\leq 12$ GeV/c} 
\vspace*{-1.0pc} 
\label{fig:xEfits}  
\end{figure*}
\section{Calculation of ${\mean{\hat{q} L}}$ from the STAR measurement~\cite{STARPLB760}.} 
\subsection{Determine $\mean{p^2_{\rm out}}$ from the $\pi^0$-h correlation function} 
\label{sec:findpout}
This is accomplished by fitting the $\pi^0$-h correlation functions for $12<p_{Tt}<20$ GeV/c~\cite{STARPLB760} to a gaussian in $\sin{\Delta\phi}$ for the away-side, $\pi/2\leq \Delta\phi\leq 3\pi/2$~\cite{ppg029,ppg089,ppg195}; and a gaussian in $\Delta\phi$ for the trigger side $-\pi/2\leq \Delta\phi\leq\pi/2$ (Fig.~\ref{fig:MJTSTARcfnfit}). The results for Au$+$Au 0-12\% centrality are $\mean{p^2_{\rm out}}=0.547\pm0.163$ (GeV/c)$^2$, $\chi^2/{\rm dof}=33/21$ for $1.2<p_{Ta}<3$ GeV/c and $\mean{p^2_{\rm out}}=0.851\pm0.203$ (GeV/c)$^2$, $\chi^2/{\rm dof}=23/21$ for $3<p_{Ta}<5$ GeV/c. The same fits to the p$+$p measurements with the same $p_{Tt}$ and $p_{Ta}$ yielded $\mean{p^2_{\rm out}}=0.263\pm0.113$ (GeV/c)$^2$, $\chi^2/{\rm dof}=186/21$ for $1.2<p_{Ta}<3$ GeV/c and $\mean{p^2_{\rm out}}=0.576\pm0.167$ (GeV/c)$^2$, $\chi^2/{\rm dof}=137/21$ for $3<p_{Ta}<5$ GeV/c where the errors have been corrected up by $\sqrt{\chi^2/{\rm dof}}$.
\subsection{Determine $\hat{x}_h=\hat{p}_{Ta}/\hat{p}_{Tt}$}
This is done by a fit of Eq.~\ref{eq:condxe2N} to the STAR measurements of what they call the away-side $z_T$ distribution~\cite{STARPLB760} (called the $x_h$ or $x_E$ distribution here) for $12<p_{Tt}<20$ GeV/c in p$+$p and Au$+$Au 0-12\% centrality collisions (Fig.~\ref{fig:xEfits}). The fit~\cite{ppg079} takes account of the statistical and correlated systematic errors, $\sigma_i$ and $\sigma_{b_i}$, for each data point with $dP/dx_E=y_i$ : 
	\begin{equation}
{\chi}^2={\left[\sum_{i=1}^{n}
{{(y_i+\epsilon_b \sigma_{b_i} -y_i^{\rm fit})^2}  \over {{\tilde{\sigma}}_i^2}} \right]}+ {\epsilon_b^2} \qquad , 
\label{eqr:lstsq}
\end{equation}
where ${\tilde{\sigma}}_{i}$ is the statistical error, $\sigma_i$, scaled by the shift in $y_i$ such that the
fractional error remains unchanged:
$\tilde{\sigma}_i=\sigma_i \left(1+\epsilon_b \sigma_{b_i}/{y_i}\right)$, where $\epsilon_b$ is to be fit.  

The fit worked very well with a result for Au$+$Au of $\hat{x}_h=0.36 \pm 0.05$ with $\chi^2/\rm{dof}=38.8/5$ where the error has been corrected upward by $\sqrt{\chi^2/\rm{dof}}$. This is consistent with the value 
$\hat{x}_h=0.48\pm 0.10$ for $9\leq p_{Tt}\leq 12$ GeV/c from a PHENIX measurement~\cite{ppg106,MJTGrenoble} (see Fig.~\ref{fig:xEfits}).

The value of $\hat{x}_h$ for the p$+$p measurement, although not needed for determining $\mean{\hat{q}L}$ in the present method, was determined for the STAR p$+$p data with fitted result $\hat{x}^{pp}_h=0.84 \pm 0.04$ 
which is in decent agreement with the result $\hat{x}^{pp}_h=0.73 \pm 0.04$ for $9\leq p_{Tt}\leq 12$ GeV/c from the PHENIX measurement (Fig.~\ref{fig:xEfits}).  
\subsection{Determine $\mean{z_t}$}
This was the easiest part of the calculation because STAR~\cite{STARPLB760} had determined that $\mean{z_t}=0.80\pm 0.05$ in their p$+$p collisions for $\pi^0$ with $12<p_{Tt}<20$ GeV/c. \vspace*{-1.0pc}
\subsection{Calculate $\mean{k_{T}^2}_{AA}$, $\mean{k{'}_{T}^2}_{pp}$, $\mean{\hat{q} L}/2$}
The $\mean{p^2_{\rm out}}$ values from the fits to the correlation functions in p$+$p and Au$+$Au plus the results for $\hat{x}^{AA}_h=0.36 \pm 0.05$, $\mean{z_t}=0.80\pm 0.05$ above are used to calculate $\sqrt{\mean{k_{T}^2}}$ using Eq.~\ref{eq:kTformula2} with the value $\sqrt{\mean{j_T}^2}=0.62\pm 0.04$ GeV/c~\cite{ppg029,ppg089} for both p$+$p and Au$+$Au. Equation~\ref{eq:qhat2} is used for $\mean{\hat{q} L}/2$.  The results are given in Table~\ref{tab:star-PLB760}.
\begin{equation}
\mean{\hat{q} L}/2=\left[\frac{\hat{x}_h}{\mean{z_t}}\right]^2 \;\left[\frac{\mean{p^2_{\rm out}}_{AA} - \mean{p^2_{\rm out}}_{pp}}{x_h^2}\right]  \qquad .
\label{eq:qhat2}
\end{equation} 
For completeness, the results for $\sqrt{\mean{k_{T}^2}}_{pp}$ with the p$+$p values $\hat{x}^{pp}_h=0.84 \pm 0.04$, $\mean{z_t}=0.80\pm 0.05$ are given in Table~\ref{tab:starpp-PLB760}. 
   \begin{table}[!h]\vspace*{-0.0pc} \label{tab:star-PLB760}
\begin{center}
\caption[]{Tabulations for $\hat{q}$--STAR $\pi^0$-h~\cite{STARPLB760}} 
{\begin{tabular}{cccc} 
\hline
STAR PLB760\\
\hline
$\sqsn=200$ &$\mean{p_{Tt}}$ & $\mean{p_{Ta}}$ & $\mean{p^2_{\rm out}}$\\ 
\hline
Reaction &   GeV/c &  GeV/c & (GeV/c)$^2$\\ 
\hline
p$+$p&14.71&1.72&$0.263\pm 0.113$\\
\hline
p$+$p&14.71&3.75&$0.576\pm 0.167$\\
\hline
\hline
Au$+$Au 0-12\%&14.71&1.72&$0.547\pm 0.163$\\
\hline
Au$+$Au 0-12\%&14.71 &3.75&$0.851\pm 0.203$\\
\hline
p$+$p comp&14.71&1.72&$0.263\pm 0.113$\\
\hline
p$+$p comp&14.71&3.75&$0.576\pm 0.167$\\ 
\hline\\[-0.9pc]
\hline
 &$\sqrt{\mean{k_{T}^2}}_{AA}$   & $\sqrt{\mean{k{'}_{T}^2}}_{pp}$    & $\mean{\hat{q} L}$\\
 \hline
Reaction & GeV/c & GeV/c &GeV$^2$ \\
 \hline
Au$+$Au 0-12\%&$2.28\pm0.35$&$1.006\pm 0.18$& $8.41\pm2.66$\\
\hline
Au$+$Au 0-12\%&$1.42\pm0.22$ &$1.076\pm 0.18$&$1.71\pm 0.67$\\
\hline\\[-1.0pc] 
\hline
\end{tabular}} \label{tab:star-PLB760}
\end{center}\vspace*{-0.1pc}
\end{table}

\begin{table}[!h]\vspace*{-0.1pc} \label{tab:starpp-PLB760}
\begin{center}
\caption[]{Tabulations for $\hat{q}$--STAR $\pi^0$-h~\cite{STARPLB760}} 
\hspace*{-0.1pc}
{\begin{tabular}{cccc} 
\hline
\hline
 & $\mean{p_{Tt}}$ & $\mean{p_{Ta}}$ & $\sqrt{\mean{k_{T}^2}}_{pp}$ \\
 \hline
Reaction & GeV/c & GeV/c &GeV/c \\
 \hline
p$+$p&14.71&1.72& $2.34\pm0.34$\\
\hline
p$+$p&14.71&3.75&$2.51\pm0.31$\\
\hline\\[-1.0pc] 
\hline
\end{tabular}} \label{tab:starpp-PLB760}
\end{center}\vspace*{-0.1pc}
\end{table}\vspace*{-1.0pc}
\section{Discussion and Conclusion}
For the $12 <p_{Tt}<20$ ($\mean{p_{Tt}}=14.71$) GeV/c, \mbox{$1.2<p_{Ta}<3$ ($\mean{p_{Ta}}=1.72$) GeV/c}  bin, the result of $\mean{\hat{q}L}=8.41\pm2.66$ GeV$^2$ agrees with the Ref.~\cite{ChenCCNU0616} result, $\mean{\hat{q}L}=14^{+42}_{-14}$ GeV$^2$,  but is not consistent with zero because of the much smaller error. The result for the $3<p_{Ta}<5$ ($\mean{p_{Ta}}=3.75$) GeV/c bin, $\mean{\hat{q}L}=1.71\pm0.67$ GeV$^2$, is at the edge of agreement, 2.4 $\sigma$ below the value in the  lower $p_{Ta}$ bin, but also 2.6 $\sigma$ from zero. If the different $p_{Ta}$ ranges do not change the original di-jet configuration, then the value of $\mean{\hat{q}L}$ should be equal in both ranges and can be weighted averaged with a result of $\mean{\hat{q}L}=2.11\pm0.64$ GeV$^2$. Taking a  guess for $\mean{L}$ in an Au$+$Au central collision as 7~fm, half the diameter of an Au nucleus, the result would be $\hat{q}=1.20\pm 0.38$ GeV$^2$/fm for the lowest $p_{Ta}$ bin, $\hat{q}=0.24\pm 0.096$ GeV$^2$/fm for the higher $p_{Ta}$ bin, with weighted average $\hat{q}=0.30\pm 0.09$ GeV$^2$/fm. These results are close to or lower than the result of the JET collaboration~\cite{JET2014} $\hat{q}=1.2\pm 0.3$ GeV$^2$/fm at $\tau_0=0.6$ fm/c. 

The new method presented here gives results for $\mean{\hat{q}L}$ comparable with the theoretical calculations noted~\cite{ChenCCNU0616,JET2014} but is more straightforward and transparent for experimentalists. This is possibly the first experimental evidence for the  predicted di-jet azimuthal broadening~\cite{BDMPSNPB484,BSZARNPS50}. It is noteworthy that the value of $\hat{x}^{AA}_h=\hat{p}_{Ta}/\hat{p}_{Tt}\approx 0.4$ combined with the 20\% loss of $\hat{p}_{Tt}$ for the trigger jet (Fig.~\ref{fig:Sloss}), which is surface biased~\cite{ppg054}, implies that the away jet has lost $\approx 3$ times more energy than the trigger jet and thus traveled a longer distance so spent a longer time in the QGP. This may affect~\cite{IvanV} the value of $\hat{q}$ used for comparison from the JET collaboration 
which used only single (trigger) hadrons for their calculation. 

It is important to emphasize that the calculated values of $\mean{\hat{q}L}$ are proportional to the square of the value of $\hat{x}_h$ derived from the measured away-side  $z_T$ (i.e. $x_E$) distribution using Eq.~\ref{eq:condxe2N}. Although in the literature for more than a decade in a well-cited paper~\cite{ppg029} and referenced in an important QCD Resource Letter~\cite{QCDAMJPhys}, Eq.~\ref{eq:condxe2N} has neither been verified nor falsified by a measurement of di-jet correlations with a di-hadron trigger. Future measurements at RHIC~\cite{MuellerPLB763,sPHENIX} will be able to do this and thus greatly improve the understanding of di-jet and di-hadron azimuthal broadening.  

\begin{acknowledgments}   
Research supported by U.~S.~Department of Energy, Contract No. {DE-SC0012704}.
\end{acknowledgments}


\begin{thebibliography}{99}
\bibitem{IvanV} Thanks to Ivan Vitev for alerting me that $\hat{q}$ calculated at a fixed $\tau_0$ will likely be larger than if integrated over the expansion or time spent in the QGP.
\end{thebibliography}
\end{document}